\documentstyle[12pt]{article}
\renewcommand{\theequation}{\thesection.\arabic{equation}}
\newcommand{\beq}{\begin{equation}}
\newcommand{\eeq}{\end{equation}}
\newcommand{\bea}{\begin{eqnarray}}
\newcommand{\eea}{\end{eqnarray}}

\begin{document}
\setcounter{page}{0}
\topmargin 0pt
\oddsidemargin 5mm
\renewcommand{\thefootnote}{\fnsymbol{footnote}}
\newpage
\setcounter{page}{0}
\begin{titlepage}

\begin{flushright}
QMW-PH-97-8\\
{\bf hep-th/9703004}\\
{\it February 1997}
\end{flushright}
\vspace{0.5cm}
\begin{center}
{\Large {\bf Deformations of a 2D charged black hole}} \\
\vspace{1.8cm}
\vspace{0.5cm}
{Oleg A. Soloviev
\footnote{e-mail: O.A.Soloviev@QMW.AC.UK}}\\
\vspace{0.5cm}
{\em
Physics Department, Queen Mary and 
Westfield College, \\
Mile End Road, London E1 4NS, United 
Kingdom}\\
\vspace{0.5cm}
\renewcommand{\thefootnote}{\arabic{footnote}}
\setcounter{footnote}{0}
\begin{abstract}
{We discuss two types of deformations of a 2D black hole carrying an 
electric charge. Type I gives rise to a space-time configuration similar to 
the one described by McGuigan, Nappi and Yost. Whereas type II results in a 
space-time configuration which has a rather peculiar geometry.  }
\end{abstract}
\vspace{0.5cm}
 \end{center}
\end{titlepage}
\newpage
\section{Introduction}

In string theory, in order to describe D-dimensional space-time 
configurations with a number of various charges, one has to consider a 
target space with extra N dimensions which accommodate corresponding 
internal degrees of freedom. Therefore, if, for example, one wants to study 
a 2D black hole carrying, say one electric charge, one has to take a string 
propagating in a three (or more) dimensional target space with the topology 
$R^2\times S^1$, where $S^1$ is a circle embedding the electric charge.  
Extra dimensions give rise to new properties of black holes. Namely, by 
virtue of extra dimensions, stringy black holes are allowed to have hair 
\cite{Larsen},\cite{Cvetic}. This fact means that a given 2D black hole 
string solution admits non-trivial perturbations along extra dimensions. It 
is argued that these dimensions can be responsible for entropy of the black 
hole as well as they can be the place where the information is stored 
\cite{Larsen},\cite{Cvetic},\cite{Horowitz}. 

Apparently, some of these deformations can be studied by means of conformal 
perturbation theory. Indeed, form the point of view of the sigma model 
approach, various deformations of string solutions are nothing but 
perturbations of given conformal sigma models. There are truly marginal and 
relevant perturbations which preserve the consistency of the non-linear 
sigma model as a two dimensional quantum field theory.

All truly marginal perturbations of a given conformal field theory, by 
definition, form the moduli space of string solutions. Changing coordinates 
in this space does not change the given conformal sigma model. At the same 
time, relevant perturbations change the CFT simply by breaking the 
conformal symmetry. However, there may be a situation when a non-conformal 
field theory flows to another critical point in the infrared limit. This IR 
CFT can be again a certain string solution. Thus relevant perturbations can 
take a string solution of one type to another string solution of a 
different type. In this case, by studying all possible relevant 
perturbations on a given string configuration, one can learn about 
dynamical properties of string theory. 

The Witten's 2D black hole is an interesting example of a non-trivial 
string solution \cite{Witten}. Therefore, it might be instructive to look 
at the effect of relevant perturbations on this 2D black hole.

In the present paper, we would like to discuss two types of relevant 
perturbations of the Witten's 2D black hole. As we shall see both these 
perturbations have one thing in common - electric field. However, their 
effects on the 2D black hole will be dramatically distinct.

\section{The basic conformal action} 

In string theory a 2D black hole without electric charge can be described 
as an $SL(2)/U(1)$ coset \cite{Witten}, which in turn is formulated as a 
gauged Wess-Zumino-Novikov-Witten model \cite{Gawedzki},\cite{Karabali}. 
The action of the gauged WZNW model is given as follows
\begin{equation}
S(g,A)=S_{WZNW}~+~{k\over2\pi}\int d^2z\mbox{Tr}\left[Ag^{-1}\bar\partial 
g-\bar A\partial gg^{-1}+Ag^{-1}\bar Ag-A\bar A\right],\end{equation}
where
\begin{equation}
S_{WZNW}={k\over8\pi}\int d^2z\mbox{Tr}g^{-1}\partial^\mu gg^{-
1}\partial_\mu g~+~{ik\over12\pi}\int d^3z\mbox{Tr}g^{-1}dg\wedge g^{-
1}dg\wedge g^{-1}dg\end{equation}
and $g\in SL(2)$, $A,~\bar A$ are the gauge fields taking values in the 
$U(1)$ algebra (compact or noncompact).

In order to be able to describe a 2D black hole carrying an electric 
charge, we have to extend the dimensionality of the target space to three. 
This can be done by adding to the action (2.1) a free scalar compactified 
on a unit circle. Then, the action of an $e=0$ (which means that there is 
no yet an electric field) 2D black hole will be given as follows
\begin{equation}
S_{3D}=S(g,A)~+~S_{S^1},\end{equation}
where
\begin{equation}
S_{S^1}={i\over4\pi}\int d^2z\bar\partial y\partial y.\end{equation}

The CFT described by eq. (2.3) is characterized by the Virasoro central 
charge
\begin{equation}
c={3k\over k+2}.\end{equation}
Formally, the given central charge coincides with the central charge of the 
ordinary WZNW model on $SL(2)$ at level $k$. As one can immediately see 
formula (2.5) differs from the central charge of the Witten's black hole
\begin{equation}
c_W={3k\over k+2}~-~1.\end{equation}
In particular, this difference results in the new value of $k$ in the 
critical case
\begin{equation}
{3k\over k+2}=26~~\longrightarrow~~ k=-{52\over23}=-2.26...,\end{equation}
whereas for the Witten's black hole
\begin{equation}
k_W=-{9\over4}=-2.25.\end{equation}

\section{$e\ne0$}

The CFT given by eq. (2.3) describes a 2D black hole without any electric 
excitations. This will be our basic conformal model.
When the electric excitations are turned on the sigma model action acquires 
a new term corresponding to the configuration of the electric field. 
Generically, this term has the following structure
\begin{equation}
S(e)=S_{3D}~-~e\int d^2z~B_\mu(x)\partial x^\mu\bar J_c,~~~~\bar 
J_c=i\bar\partial y,\end{equation}
and $B_\mu$ is the vector gauge potential, a function of $x^\mu$ only, 
with $x^\mu$ being the coordinates of the 2D target space, and $e$ is 
proportional to the electric charge. Admissible configurations of $B_\mu$ 
are fixed by the conformal invariance of the non-linear sigma model in eq. 
(3.9). 

If we choose $e$ to be a small parameter, then one can treat model (3.9) 
within perturbative approach. In this case, one has to consider all 
possible relevant deformations of the form (3.9) constructed in terms of 
the CFT $S_{3D}$. A natural choice appears to be as follows
\begin{equation}
B_\mu\partial x^\mu\bar J_c\equiv O_B(z,\bar 
z)=L_{ab}J^a(z)\phi^{b\bar3}(z,\bar z)\bar J_c,\end{equation}
where
\begin{equation}
J=-{k\over2}\partial gg^{-1},~~~~~~\phi^{a\bar a}=\mbox{Tr}(t^agt^{\bar 
a}g^{-1}),\end{equation}
and
\begin{equation}
L={1\over A}\left(\begin{array}{ccc}
1&0&0\\
0&1&0\\
0&0&-{4\over k}\\
\end{array}\right).\end{equation}
where $A$ is a normalization constant. The form of the matrix $L$ is fixed 
by the invariance of the operator $O_B$ under the gauge group $U(1)$.

The important point to be made is that the conformal dimension of the 
operator $O_B$ is
\begin{equation}
\Delta=1~+~{2\over k+2},\end{equation}
that is for all $k<-2$, $O_B$ is a relevant operator. Moreover, for all 
$k<-4$, this is a relevant operator with positive conformal dimension. In 
what follows, we shall be interested in the limit $k\to-\infty$. In this 
case, $O_B$ is a relevant quasimarginal operator. From the point of view of 
the string criticality condition, this limit means that we consider an 
electrically charged 2D black hole with 24 additional internal dimensions 
one of which accommodates the electric charge, whereas the other 23 charges 
are set to zero. In total, the dimensionality of the whole target space is 
equal to 26. Thus, we are dealing with a certain solution of the critical 
bosonic string.

Now we are going to study possible relevant perturbations of type (3.9). It 
turns out that perturbation theory allows two different types of 
deformations of the space-time configuration of the charged 2D black 
hole(s).

\subsection{\it $e\ne0$ type I}

By adding the relevant operator $O_B$ to the CFT $S_{3D}$ we do not 
automatically obtain a renormalizable quantum field theory. Indeed, the 
operator $O_B$ has the following operator product expansion (see appendix)
\begin{equation}
O_B(z,\bar z)O_B(0)={A^2\over2(1+k/2)|z|^{2\Delta}}O(0)~+~{\Large\rm 
I}~+~...,\end{equation}
where
\begin{equation}
O(z,\bar z)=L_{ab}L_{\bar a\bar b}J^a\bar J^{\bar a}\phi^{a\bar a}
\end{equation}
is a new operator with the same conformal dimension as in eq. (3.13), I is 
the unity operator and dots stand for all other operators with conformal 
dimensions greater than 1. Thus, the CFT $S_{3D}$ perturbed by $O_B$ alone 
is not renormalizable and one has to add also the operator $O$ with the 
corresponding coupling constant. From the physical point of view, this 
operator corresponds to the back reaction of the 2D metric to the electric 
charge. Then, the perturbed theory takes the following form
\begin{equation}
S_I(e,\epsilon)=S_{3D}~-~\epsilon\int d^2z O(z,\bar z)~-~e\int 
d^2zO_B(z,\bar z).\end{equation}
The given QFT is renormalizable by virtue of the fact that the operators 
$O,~O_B$ form a closed algebra with respect to their OPE's:
\begin{eqnarray}
O(z,\bar z)O(0)&=&{1\over|z|^{2\Delta}}O(0)~+~{\Large\rm I}~+~...,\nonumber 
\\ & & \\
O(z,\bar 
z)O_B(0)&=&{1\over|z|^{2\Delta}}O_B(0)~+~...\nonumber\end{eqnarray}

We shall consider the theory in eq. (3.16) as type I. Its characteristic 
features are that the perturbation (3.16) does not excite the metric 
component of the hidden dimension, however, it generates a non-trivial WZ 
term. Indeed, the operator $O_B$ is not symmetric in $\partial$ and 
$\bar\partial$. 

Given the OPE's (3.14), (3.17), it is not difficult to calculate the 
corresponding renormalization group beta functions of $e$ and $\epsilon$. 
We find
\begin{eqnarray}
\beta_\epsilon&=&(2-2\Delta)\epsilon~-~\pi\epsilon^2~-~{\pi A^2\over 
k}e^2~+~...,\nonumber\\ & & \\
\beta_e&=&(2-2\Delta)e~-~\pi\epsilon e~+~...\nonumber\end{eqnarray}
If we assume that $e$ is very small, then one can ignore the $e^2$ 
contribution to $\beta_\epsilon$ in the limit $k\rightarrow-\infty$. In 
this limit, one can easily see that the equations (3.18) admit a non-
trivial infrared conformal point at
\begin{equation}
\epsilon^\star={2-2\Delta\over\pi}~+~...=-{4\over\pi k}~+~{\cal 
O}(1/k^2).\end{equation}
Note that when $\epsilon$ is set to its critical value $\epsilon^\star$, 
the charge $e$ gets no restriction on its value apart from the requirement 
to be small. In other words, at the point $\epsilon^\star$, the perturbed 
theory(3.16) becomes a CFT with a continuous parameter $e$. 
Correspondingly, the Virasoro central charge at the IR critical point is 
given as follows (see appendix)
\begin{equation}
c(\epsilon^\star)={3k\over k+2}~+~{256\over3 k}~+~{\cal 
O}(1/k^2).\end{equation}
Unfortunately, the given perturbative value of the Virasoro central charge 
does not appear to fit any known exact CFT. However, from the string theory 
point of view, at the IR fixed point, the perturbed CFT (3.16) describes a 
2D electrically charged black hole whose metric looks similarly to the 
metric found in \cite{Guigan}.

\subsection{\it $e\ne0$ type II}

There exists another deformation of the CFT $S_{3D}$ which differs from the 
perturbation described above. Let us consider the following operator
\begin{equation}
O_+(z,\bar z)=\left({1\over2}L_{ab}J^a(z)\phi^b(z)~+~x\phi^3(z)J_c(z)
\right)\overline{\left({1\over2}L_{ab}J^a(z)\phi^b(z)~+~x\phi^3(z)J_c(z)
\right)},
\end{equation}
where the parameter $x$ is defined from the condition
\begin{equation}
O_+(z,\bar z)O_+(0)={1\over |z|^{2\Delta}}O_+(0)~+~{\Large\rm 
I}~+~...\end{equation}
This condition is satisfied when
\begin{equation}
x={\pm i\sqrt{|k|}\over2A}.\end{equation}
The operator $O_+$ has the conformal dimension as in eq. (3.13).

Now we can consider the following deformation
\begin{equation}
S_{II}(\mu)=S_{3D}~-~\mu\int d^2zO_+(z,\bar z),\end{equation}
where $\mu$ is a new coupling constant. It is not difficult to see that 
this perturbed CFT includes type I perturbation with
\begin{equation}
e=\mu x/2.\end{equation}
However, compared to the model in eq. (3.16), the theory (3.24) introduces 
another electric field $\bar B_\mu$ associated with the string modes of the 
opposite chirality in the compact dimension. Additionally, the operator 
$O_+$ gives rise to the excitation of the metric of the hidden dimension. 
Thus, the new perturbed CFT describes a few more processes occurring in the 
compact dimension. This can be viewed as a further deformation of the type 
I perturbation by extra operators
\begin{equation}
O_{\bar B}\equiv\bar B_\mu J_c\bar\partial x^\mu=L_{\bar a\bar b}\bar 
J^{\bar a}\phi^{3\bar b}J_c,~~~~~~O_{3\bar 3}=\phi^{3\bar 3}J_c\bar 
J_c.\end{equation}

Because $O_+$ forms a closed OPE algebra, the perturbed CFT (3.24) is a 
renormalizable QFT and one can compute the corresponding beta function. We 
find
\begin{equation}
\beta_\mu=(2-2\Delta)\mu~-~\pi\mu^2~+~...\end{equation}
This equation also exhibits a non-trivial IR fixed point at
\begin{equation}
\mu^\star=-{4\over\pi k}~+~{\cal O}(1/k^2).\end{equation}
Interestingly, at this value of $\mu$, the electric charge $e$ is fixed at 
the value corresponding to the so-called extremal 2D black hole. Namely,
\begin{equation}
e={\pm 4i\over \pi A\sqrt{|k|}}.\end{equation}

At the IR conformal point, the most remarkable thing happens to the 
Virasoro central charge. At this point, it is given by the following 
perturbative formula (see appendix)
\begin{equation}
c(\mu^\star)={3k\over k+2}~+~{12\over k}~+~{\cal O}(1/k^2).\end{equation}
This expression needs to be carefully examined before we can identify the 
IR conformal point with any exact CFT. There are two candidates which fit 
the perturbative expansion (3.30). Namely, the $SU(2)_{|k|}$ WZNW model and 
the $(SU(2)_{|k|}/U(1))\times U(1)$ coset construction. Both the WZNW model 
and the coset construction have one and the same Virasoro central charge, 
which in the large $|k|$ limit has the form given by eq. (3.30). In order 
for the IR CFT to be the $SU(2)$ WZNW model, it has to have the 
corresponding affine currents. However, the perturbed theory does not 
appear to have affine currents which form the affine algebra 
$\widehat{SU(2)}$. At the same time, the perturbed theory (3.24) still has 
the BRST symmetry, since the perturbation operator is BRST invariant. Thus, 
at the IR conformal point one has to expect to arrive at a certain gauge 
invariant model whose gauge group has to be $U(1)$. Apparently, the 
$(SU(2)_{|k|}/U(1))\times U(1)$ coset has all these features, provided that 
we have gauged away the compact subgroup of $SL(2)$ at the UV fixed point. 
Hence, we come to conclusion that at the IR critical point, the perturbed 
CFT coincides with $(SU(2)_{|k|}/U(1))\times U(1)$ in the large $|k|$ 
limit. All in all, we can have the following picture of the corresponding 
renormalization group flow
\begin{equation}
{SL(2)_k\over U(1)}\times U(1)\longrightarrow {SU(2)_{|k|}\over U(1)}\times 
U(1).\end{equation}
The given flow describes the change the target space geometry undergoes 
under the perturbation by the operator $O_+$. Indeed, the UV conformal 
point of the flow corresponds to the Witten's black hole without electric 
charge, whose Euclidean target space geometry is presented as a cigar. 
After adding the perturbation of type II, we observe that along this flow 
the 2D black hole settles down to the configuration described by the coset 
$(SU(2)_{|k|}/U(1))\times U(1)$, in the limit $k\to-\infty$. Naively, it 
may seem that the given coset describes a charged two dimensional sphear. 
However, Bardakci, Crescimanno and Rabinovici have shown \cite{Bardakci} 
(see also \cite{Bakas}) that $SU(2)/U(1)\ne S^2$. In fact this coset has a 
rather peculiar geometry of a 2D sphear with blown up equator (or two 
Mexican hats with infinitely large brim glued together at the rim). Also it 
describes parafermions \cite{Fateev}. Unfortunately, not very much is known 
about string solutions with this type target-space geometry.

\section{Conclusion}

We have studied two types of perturbations of the Witten's 2D black hole. 
These perturbations have one thing in common - they excite an electric 
field. However, the difference between these two perturbations is in their 
fate as they approach the IR limit. Type I perturbation evolves into a 
configuration, which has an interpretation of the 2D charged black hole. 
Whereas type II perturbation, at the IR conformal point ceases to look like 
a 2D black hole, but acquires a rather peculiar target-space geometry.

We also would like to point out that we have not discussed how the 
considered perturbations affect the dilaton field. This may be important in 
order to understand whether these two perturbations are related to each 
other in any way. 

\section{Acknowledgments}

I am indebted to Ian Kogan for very significant discussions.

\bigskip
\bigskip
{\large\bf Appendix }\vspace{.15in}
\renewcommand{\theequation}{A.\arabic{equation}}
\setcounter{equation}{0}

In this appendix we would like to explain how OPE's (3.14), (3.17) come 
into being. First of all, let us take the holomorphic operator
\begin{equation}
O(z)\equiv L_{ab}J^a(z)\phi^b(z),\end{equation}
where $\phi^a(z)$ is the holomorphic part of $\phi^{a\bar a}(z,\bar z)$. 
One can demand that $O(z)$ obeys the following OPE
\begin{equation}
O(z)O(0)={1\over z^\Delta}O(0)~+~{\Large\rm I}~+~...\end{equation}
This requirement results in a condition on the matrix $L_{ab}$ 
\cite{Soloviev1}
\begin{eqnarray}
&&2\left({k\over2}g^{k(m}\delta^{n)}_p~-
~f^{k(m}_ff^{n)f}_p\right)L_{mn}=L_{ab}L_{cd}\left\{{k\over2}\left[g^{ka}C^
{cd,b}_p~+~g^{kb}C^{cd,a}_p\right.\right.\nonumber\\ & & \\
&&\left.\left.+~g^{kc}C^{ab,d}_p~+~g^{kd}C^{ab,a}_p\right]~-~
f^{k(a}_ff^{b)f}_eC^{cd,e}_p~-~f^{k(c}_ff^{d)f}_eC^{ab,c}_p~+~
f^{k(a}_eC^{b)e,cd}_p \right\},\nonumber\end{eqnarray}
with
\begin{eqnarray}
C^{ab,c}_d&=&{A\over2}\left(f^{bc}_ef^{ea}_d~+~f^{ac}_ef^{eb}_d\right),
\nonumber\\ & & \\
C^{ab,cd}_e&=&{A\over4}\left[\left(f^{af}_ef^{bd}_nf^{cn}_f~+~f^{af}_ef^{bc
}_nf^{dn}_f~+~f^{cf}_ef^{ad}_nf^{bn}_f~+~f^{df}_ef^{ac}_nf^{bn}_f\right)~+~
(a\leftrightarrow b)\right].\nonumber\end{eqnarray}

The matrix
\begin{equation}
L={1\over A}\left(\begin{array}{ccc}
1&0&0\\
0&1&0\\
0&0&-{4\over k}\\
\end{array}\right)\end{equation}
is one of the solutions of eq. (A.3). The constant $A$ can be computed from 
the normalization condition which stems from the classical identity
\begin{equation}
J^a=\phi^{a\bar a}\phi^{b\bar a}J^b.\end{equation}
This means that in the large $k$ limit
\begin{equation}
\phi^{a\bar a}\phi^{b\bar a}=g^{ab}.\end{equation}
Finally, we find
\begin{equation}
A^2=1/c_V,~~~~~c_Vg^{ab}=-f^{ac}_df^{bd}_c.\end{equation}
In the case under consideration, $c_V=2$.

Alternatively, the matrix $L_{ab}$ can be derived from the BRST condition
\begin{equation}
QO(0)|0\rangle=0,\end{equation}
where the BRST charge $Q$ is defined as follows
\begin{equation}
Q=\oint{dw\over2\pi i}:c(\tilde J^3~+~J^3):(z).\end{equation}
Here $c$ is the $U(1)$ ghost and $\tilde J^3$ is the current associated 
with the subgroup $H=U(1)$,
\begin{equation}
\tilde J^3(z)\tilde J^3(0)={-k/2\over w^2}~+~{\rm reg.terms}.\end{equation}

Since $\phi^3$ is also a BRST invariant operator, all OPE's involving $O$ 
and $\phi^3$ must be BRST invariant. This symmetry principle allows us to 
determine which operators may appear on the right hand side of OPE's. Then 
the exact OPE coefficients can be calculated from the consistency 
conditions \cite{Soloviev1}. So we find
\begin{equation}
\phi^3(z)\phi^3(0)={A^2\over(1+k/2)z^{\Delta_\phi-1}}O(0)~+~{\Large\rm 
I}~+~...\end{equation}
Now it is easy to see that
\begin{equation}
\phi^{\bar 3}(\bar z)\bar J_c(\bar z)\phi^{\bar 3}(0)\bar J_c(0)\bar 
={A^2\over(2+k)\bar z^\Delta}\bar O(0)~+~{\Large\rm I}~+~...\end{equation}
Together with (A.2) this gives rise to the OPE in eq. (3.14).

Similarly, the OPE of $O$ with $O_B$ follows from the formula
\begin{equation}
\phi^c(z)O(0)={L_{ab}C^{ab,c}_d\over2z^{\Delta_\phi}}\phi^d(0)~+~...,
\end{equation}
where $C^{ab,c}_d$ is given by (A.4).

Let us also show how we compute perturbative Virasoro charges (3.20), 
(3.30). According to Zamolodchikov's c-theorem \cite{Zamolodchikov}, the 
Virasoro central charge at the IR fixed point is given by the following 
formula \cite{Cardy}
\begin{equation}
c(IR)=c(UV)~-~{(2-2\Delta_O)^3||O||^2\over C^3}~+~...,\end{equation}
where $||O||$ is the norm of the perturbation operator and $C$ is the 
coefficient in the following OPE
\begin{equation}
O(z,\bar z)O(0)={C\over|z|^{2\Delta_O}}O(0)~+~....
\end{equation}

In type I perturbation, in the limit $k\to-\infty$, only the operator $O$ 
contributes into leading order corrections at the IR conformal point. We 
find
\begin{equation}
C_I=1,~~~~~~||O||^2={k^2\sum^2_{i,\bar i=1}||\phi^{i\bar i}||^2\over 
4A^4}~+~....\end{equation}
Here the norm of the operator $\phi$ can be derived from the normalization 
condition (A.7) \cite{Soloviev2},
\begin{equation}
||\phi^{a\bar a}\phi^{b\bar b}||={\delta^{ab}\delta^{\bar a\bar b}\over\dim 
G}~+~{\cal O}(1/k).\end{equation}
Thus,
\begin{equation}
c(IR)_I=c(UV)~+~{256\over3k}~+~{\cal O}(1/k^2).\end{equation}

For type II we find
\begin{equation}
C_{II}=1,~~~~~~||O_+||^2={k^2\sum^3_{a\bar a=1}||\phi^{a\bar a}||^2\over 
4^3A^4}~+~....\end{equation}
Correspondingly,
\begin{equation}
c(IR)_{II}=c(UV)~+~{12\over k}~+~{\cal O}(1/k^2).\end{equation}

\end{document}